\documentclass[aps,prb,showpacs,floatfix,twocolumn,byrevtex,superscriptaddress]{revtex4-1}
\usepackage{amsmath}
\usepackage{amssymb}
\usepackage{amstext}
\usepackage{amsopn}
\usepackage{amsfonts}
\usepackage{amsxtra}
\usepackage[english]{babel}
\usepackage{amsmath}
\usepackage{graphicx}
\usepackage{float}
\usepackage{bm}
\usepackage{multirow}
\usepackage{dcolumn}
\usepackage{hyperref}
\usepackage{CJK}

\newcommand{\icm}{\ensuremath{~\textrm{cm}^{-1}}}
\newcommand\cms{CaMnSb$_{2}$}
\newcommand{\cmb}{CaMnBi$_{2}$}
\newcommand{\ams}{$A$MnSb$_{2}$}
\newcommand{\amb}{$A$MnBi$_{2}$}

\newcommand{\nb}{Na$_{3}$Bi}
\newcommand{\ca}{Cd$_{3}$As$_{2}$}
\newcommand{\smb}{SrMnBi$_{2}$}
\newcommand{\bms}{BaMnSb$_{2}$}

\newcommand{\sms}{SrMnSb$_{2}$}
\newcommand{\ymb}{YbMnBi$_{2}$}
\newcommand{\emb}{EuMnBi$_{2}$}
\newcommand{\cfa}{CaFeAs$_{2}$}

\begin{document}
\title{Infrared spectroscopic studies of the topological properties in CaMnSb$_{2}$}
\author{Ziyang Qiu}
\affiliation{Beijing National Laboratory for Condensed Matter Physics, Institute of Physics, Chinese Academy of Sciences, P.O. Box 603, Beijing 100190, China}
\affiliation{School of Physical Sciences, University of Chinese Academy of Sciences, Beijing 100049, China}
\author{Congcong Le}
\affiliation{Beijing National Laboratory for Condensed Matter Physics, Institute of Physics, Chinese Academy of Sciences, P.O. Box 603, Beijing 100190, China}
\affiliation{Kavli Institute of Theoretical Sciences, University of Chinese Academy of Sciences, Beijing, 100049, China}
\author{Yaomin Dai}
\affiliation{Center for Superconducting Physics and Materials, National Laboratory of Solid State Microstructures and Department of Physics, Nanjing University, Nanjing 210093, China}
\author{Bing Xu}
\affiliation{Beijing National Laboratory for Condensed Matter Physics, Institute of Physics, Chinese Academy of Sciences, P.O. Box 603, Beijing 100190, China}
\author{J. B. He}
\affiliation{Beijing National Laboratory for Condensed Matter Physics, Institute of Physics, Chinese Academy of Sciences, P.O. Box 603, Beijing 100190, China}
\author{Run Yang}
\affiliation{Beijing National Laboratory for Condensed Matter Physics, Institute of Physics, Chinese Academy of Sciences, P.O. Box 603, Beijing 100190, China}
\affiliation{School of Physical Sciences, University of Chinese Academy of Sciences, Beijing 100049, China}
\author{Genfu Chen}
\affiliation{Beijing National Laboratory for Condensed Matter Physics, Institute of Physics, Chinese Academy of Sciences, P.O. Box 603, Beijing 100190, China}
\affiliation{School of Physical Sciences, University of Chinese Academy of Sciences, Beijing 100049, China}
\affiliation{Collaborative Innovation Center of Quantum Matter, Beijing 100084, China}
\author{Jiangping Hu}
\affiliation{Beijing National Laboratory for Condensed Matter Physics, Institute of Physics, Chinese Academy of Sciences, P.O. Box 603, Beijing 100190, China}
\affiliation{Kavli Institute of Theoretical Sciences, University of Chinese Academy of Sciences, Beijing, 100049, China}
\affiliation{Collaborative Innovation Center of Quantum Matter, Beijing 100084, China}
\author{Xianggang Qiu}
\email[]{xgqiu@iphy.ac.cn}
\affiliation{Beijing National Laboratory for Condensed Matter Physics, Institute of Physics, Chinese Academy of Sciences, P.O. Box 603, Beijing 100190, China}
\affiliation{School of Physical Sciences, University of Chinese Academy of Sciences, Beijing 100049, China}
\affiliation{Collaborative Innovation Center of Quantum Matter, Beijing 100084, China}
%
%

\begin{abstract}
We present temperature-dependent infrared spectroscopic studies of \cms, a proposed three-dimensional topological material. The low plasma edge in the reflectivity spectrum and small Drude component in the optical conductivity indicate a very low carrier density. The low-frequency optical conductivity is well described by the superposition of a narrow and a broad Drude terms. Several linear components have been observed in the low-temperature optical conductivity, but none of them extrapolates to the origin, at odds with the optical response expected for three-dimensional Dirac fermions. A series of absorption peaks have been resolved in the high-frequency optical conductivity. The energy of these peaks agrees well with the interband transitions expected for the band structures from first-principles calculations. Intriguingly, the lowest band gap increases with decreasing temperature, mimic the temperature evolution of inverted bands. Furthermore, our theoretical calculations demonstrate the existence of weak coupling between two Sb-chains layers results in the topological trivial surface states in \cms.
\end{abstract}


\maketitle

%

\section{Introduction}
Topological materials, such as topological insulators (Pb$_{1-x}$Sn$_{x}$Se, Bi$_{2}$Se$_{3}$, Bi$_{2}$Te$_{3}$ and Sb$_{2}$Te$_{3}$~\cite{Xi2014,Dziawa2012,Zhang2009}), Dirac (\nb\ and\ \ca\cite{wang2012, wang2013,Akrap2016}) and Weyl semimetals (TaAs and TaP~\cite{weng2015, Huang2015}), have been extensively explored theoretically and experimentally. A three-dimensional (3D) topological insulator with gapped bulk states can support odd numbers of gapless Dirac nodes in its surface states. For a Dirac semimetal, valance and conduction bands linearly touch near the Fermi level, resulting in a Dirac cone-like electronic dispersion in the 3D momentum space. The band touching point is called a Dirac point. When either time-reversal or inversion symmetry is broken, a Dirac point is split into two Weyl points with opposite chiralities, turning a Dirac semimetal into a Weyl semimetal(WSM). Since topological materials exhibit rich quantum phenomena, such as quantum Hall effect, topological magnetoelectric effect, Fermi arc, as well as Majorana excitations~\cite{Bansil2016,Hasan2010,Qi2011}, these materials have aroused an enormous amount of interests.

Recently, layered pnictide Bi/Sb compounds, such as \amb\ and \ams\ (A = Sr, Ba, Eu, and Yb), are proposed to be Dirac or Weyl semimetals with magnetic order on the Mn sublattice through both theoretical and experimental investigations~\cite{Park2011,wangcmb2012,Li2016,wangymb2016,wangsmb2011,Andrew2014, Chinotti2016,Lee2013,Liu2016,Chaudhuri2017}. They have a crossing of two non-degenerate energy bands, containing relativistic and massless states that touch in a single point, which must be satisfied for the search of three-dimensional Dirac fermions. Among \amb\ family, \smb, \cmb\ and \emb\ have been proved as topological semimetals with their band structures protected by space group symmetry~\cite{Chinotti2016,Park2011,He2012}. Nevertheless, for \ymb , even though gapless Weyl points have been observed by ARPES~\cite{Borisenko2015}, the optical studies speculate that the evidence for a WSM observed in ARPES could possibly arise through a surface magnetic phase~\cite{Chaudhuri2017}. Since the \ams\ system with weaker spin-orbit coupling (SOC) and massless Dirac fermions~\cite{Farhan2014} is so far relatively unexplored, detailed studies on this system may shed new light on the topological properties of the layered pnictide Bi/Sb compounds.

As a member of the \ams\ family, \sms\ has an orthorhombic \emph{Pnma} structure with a staggered stacking of Sr atoms and zig-zag chains of Sb atoms, which is quite different from the structure with coincident stacking in \amb\ and \bms\cite{Liu2015,BRECHTEL1981131}. In \sms\, even though the Fermi surface mapped by ARPES agrees well with the de Haas van Alphen (dHvA) data, discrepancies exist between theoretical calculations and ARPES results~\cite{Ramankutty2017}. In order to resolve these discrepancies, it is of great importance to conduct a comprehensive study on the \ams\ family. \cms, which is isostructural with \sms\ ~\cite{Farhan2014}, has been reported to host massless Dirac fermions from dHvA oscillation measurements in Ref.\cite{He2017}. However, there are few reports on the topological properties in \cms, so it is of great interest to further investigat the topological properties of \cms.

Infrared spectroscopy is a powerful tool to study the excitations near the Fermi level which has been previously used to investigate properties related to the band structures in topological materials~\cite{Hosur2012, Ashby2014,Aemitage2018}. In this paper, we report the observation of several linear components in the frequency-dependent optical conductivity associated with the linear dispersions in the electronic band structure. However, neither of them extrapolates to the origin, in conflict with the expected behavior for a 3D Dirac semimetal. In combination with theoretical calculations, we conclude that \cms\ is a topological trivial insulator, and the non-robust gapless surface states can be identified when the coupling between two Sb-chains layers is ignored.

%
%
\section{Experiment}

High-quality single crystal of \cms\ has been synthesized using a self-flux method~\cite{He2017}, which shows an antiferromagnetic (AFM) phase transition at 302~K. The frequency-dependent reflectivity $R(\omega)$ has been measured at a near-normal angle of incidence on a Bruker 80v Fourier transform infrared (FTIR) spectrometer using an \emph{in situ} evaporation technique~\cite{Homes1993}. Data from 50 to 15\,000 \icm\ were collected from 5 to 315~K at 15 different temperatures with the sample mounted in an ARS-Helitran cryostat. $R(\omega)$ in the visible and UV range (10\,000-30\,000~\icm) was taken at room temperature with an Avaspec
$2048 \times 14$ optical fiber spectrometer. The real part of the optical conductivity $\sigma_{1}(\omega)$ has been obtained from a Kramers-Kronig transformation of $R(\omega)$. Usually, for the low-frequency ($< 50~\icm$) extrapolation, Hagen-Rubens form ($R=1-A\sqrt{\omega}$) is employed. However, in our case, the Drude component is very narrow with a scattering rate about 20~\icm, which is much smaller than $\nu_{min}\approx100~\icm$. Thus a set of Lorentzians are used to extrapolated the measured reflectivity~\cite{Schilling2017,Chaudhuri2017}. Above the highest-measured frequency (30\,000~\icm), $R(\omega)$ is assumed to be constant up to 12.4 eV, above which a free-electron response ($\omega^{-4}$) is used.

\section{Results}
\subsection{Reflectivity and optical conductivity}
%
%

\begin{figure}
  \includegraphics[width=0.9\columnwidth]{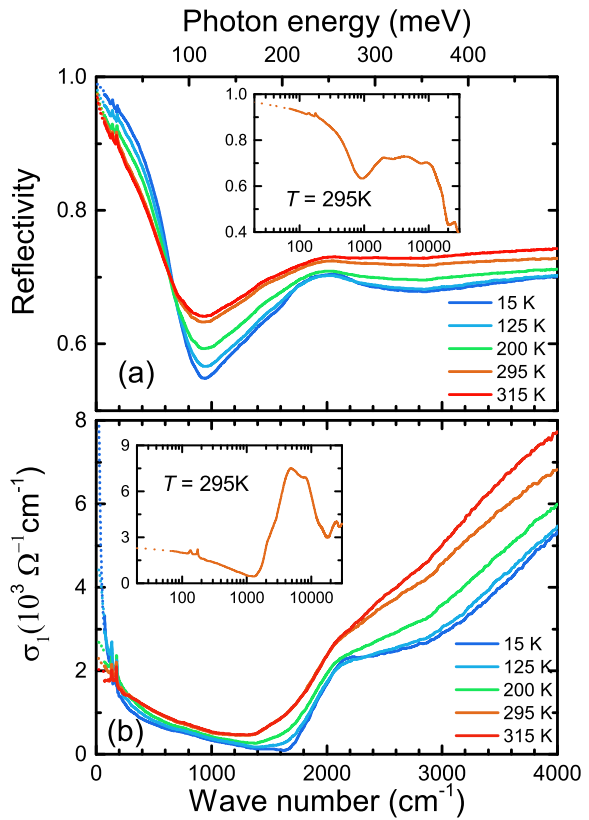}\\
  \caption{(Color online) Reflectivity (a) and the real part $\sigma_1(\omega)$ of the optical conductivity (b) of \cms\ at 5 representative temperatures. The insets of the main panels show the reflectivity and the optical conductivity over a broad frequency range at 295~K, respectively. The color dashed lines are the sets of Lorentzians for extrapolation.}
  \label{sigma}
\end{figure}

Fig.~\ref{sigma}(a) shows the measured $R(\omega)$ of \cms\ up to 4\,000~\icm\ at several representative temperatures. The sharp features around 200~\icm\ are associated with the infrared-active phonons. The low plasma edge (500~\icm) in $R(\omega)$ indicates a low carrier density. Fig.~\ref{sigma}(b) displays $\sigma_{1}(\omega)$ of \cms\ at different temperatures. The low-frequency $\sigma_{1}(\omega)$ ($< 1\,500$~\icm) is dominated by the Drude-like response. However, the Drude weight (the area under the Drude peak in the optical conductivity spectrum) in this material is remarkably smaller than that of a normal metal~\cite{Beach1977}, consistent with the low plasma edge in $R(\omega)$. Above 1\,500~\icm, $\sigma_{1}(\omega)$ increases sharply, leading to a noticeable absorption edge, which is a signature of the emergence of inter-band transitions. Multiple absorption peaks arising from the inter-band transitions can be identified in the high-frequency $\sigma_{1}(\omega)$, as shown in the inset of Fig.~\ref{sigma}(b). The overall optical conductivity exhibits a strongly-frequency dependence, which is in sharp contrast to the frequency-independent conductivity observed in two-dimensional(2D) Dirac semimetals\cite{Bacsi2013,Schilling2017-2}.

\subsection{Free carrier response}
\begin{figure}
  \centering
  \includegraphics[width=0.9\columnwidth]{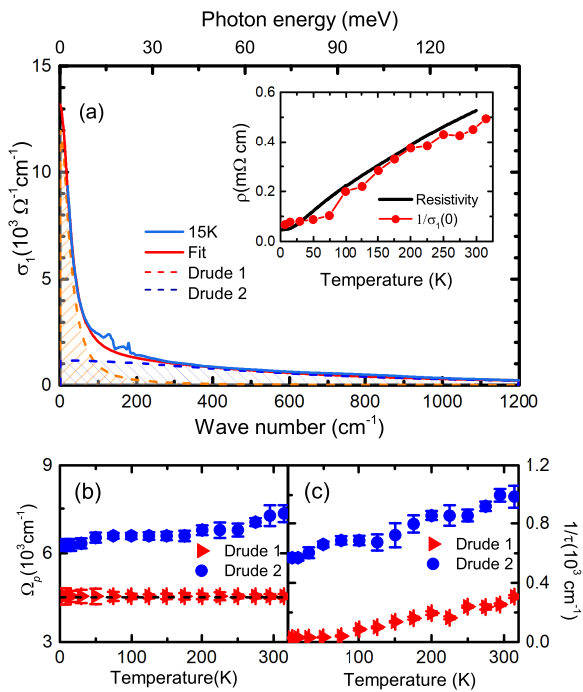}\\
  \caption{(Color online). (a) Fit (orange dashed curve) to the optical conductivity (cyan solid curve) in the frequency range of 0-1200~\icm\ at 15~K, which is decomposed into two Drude components: a narrow (red solid curve) and a broad one (blue solid curve). Inset: a comparison between the temperature-dependent dc resistivity (solid curve) from transport measurements and the values derived from the zero-frequency extrapolation of $\sigma_1(\omega)$ (solid circles). The temperature dependence of (b) the plasma frequency $\Omega_\emph{p}$ and (c) the scattering rate $1/\tau$ for the narrow (red) and broad (blue) Drude components. }
  \label{fit}
\end{figure}

In order to quantitatively analyze the optical data, we fit the low-frequency $\sigma_1(\omega)$ (below 1\,200~\icm) to the Drude model, which is used to describe the optical response of free carriers~\cite{Dai2014}:
\begin{equation}
\label{DrudeLorentz}
\sigma_1(\omega)=\frac{2\pi}{Z_0}\sum_k\frac{\Omega_{p,k}^2}{\tau_k(\omega^2+\tau_k^{-2})},
\end{equation}
where $Z_0 = 377~\Omega$ is the impedance of the free space; $\Omega_p=\sqrt{4\pi ne^2/m^*}$ the plasma frequency ($n$ is a carrier density and $m^*$ is an effective mass); $1/\tau$ the scattering rate of carriers.

We find that a single Drude term fails to describe the low-frequency $\sigma_{1}(\omega)$, because of a pronounced background feature extending up to 1\,500~\icm, far beyond the Drude response we adopted. We have tried to fit the conductivity with an additional frequency-independent component which is a feature associated with 2D Dirac material, but it turns out that one Drude component plus one frequency-independent term is not possible to make the fitting at different temperatures. Therefore, a second broad Drude component has to be introduced to account for the background feature. Using this two-Drude model, we can fit $\sigma_{1}(\omega)$ at all measured temperatures. The solid circles in the inset of Fig.~\ref{fit}(a) give the zero-frequency conductivity of the Drude model fit, which is compared to the dc resistivity from transport measurements (the solid curve). The reasonable agreement between the optical and transport data implies that our modeling is reliable. The two Drude components usually point to the existence of multiple Fermi surfaces. Recent quantum oscillation and Hall effect studies have revealed multi-band Fermi surfaces with very small cross-sectional areas in this material~\cite{He2017}, which is in agreement with our observations.

The two-Drude model fit yields the temperature dependence of $\Omega_{p}$ and $1/\tau$ for each component. Figure~\ref{fit}(b) depicts $\Omega_p$ as a function of the temperature for the two Drude components. Within error bar, $\Omega_{p}^{\prime}$s of Drude 1 exhibit no temperature dependence, indicating that the carrier densities or Fermi surface volumes is not affected by the temperature, while the $\Omega_{p}^{\prime}$s of Drude 2 increases slowly with the increasing temperature. The temperature evolution of the quasi-particle scattering rate is traced out in Fig.~\ref{fit}(c). $1/\tau$ decreases upon cooling for both Drude terms, giving rise to a continuous drop of the dc resistivity. This suggests that the transport properties of \cms\ are dominated by the change of quasi-particle scattering rate.

\begin{figure}
  \includegraphics[width=0.9\columnwidth]{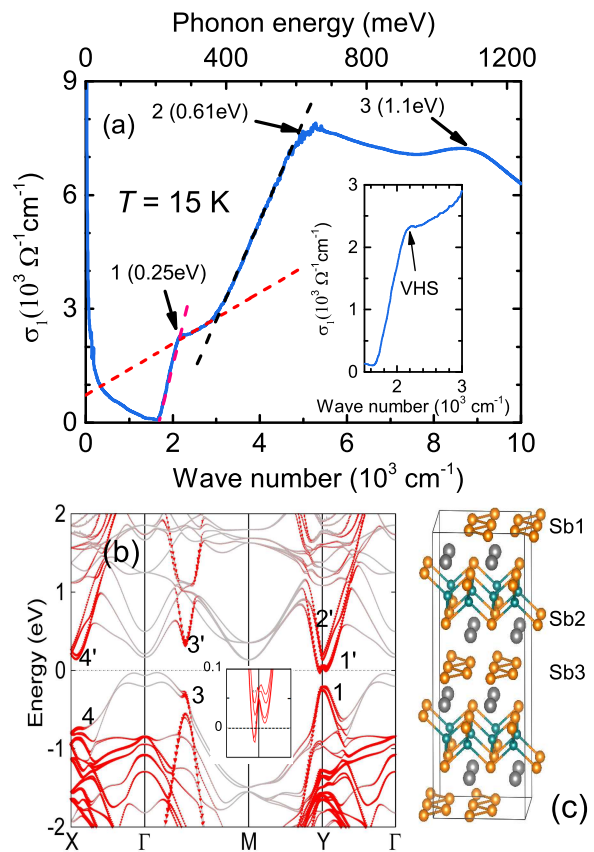}\\
  \caption{(Color online) (a) Optical conductivity in the high-frequency range showing interband transitions for \cms\ at 15K (cyan curve). The pink dashed line is the linear fit to the sharp absorption edge. The red and black dashed lines through the data denote linear fits. The inset is the van-Hove singularity risen from the Mexican-hat dispersion. (b) Calculated band structure of \cms. The inset displays the enlarge view of the band structure near the $Y$ point. (c) The crystal structure of \cms.}
  \label{interband}
\end{figure}

\subsection{Interband transitions}

Figure~\ref{interband}(a) shows $\sigma_{1}(\omega)$ at 15~K up to 10\,000~\icm. Besides a linear rise of the sharp absorption edge depicted by the pink dashed line, another two $\omega$-linear components with distinct slopes can be clearly identified in the optical conductivity spectrum as denoted by the red and black dashed lines. Linear optical conductivity is usually associated with the linear dispersion in the electronic structures, which has often been observed in 3D Dirac/Weyl semi-metals~\cite{Neubauer2016, chen2015, Chinotti2016, Xu2016} or Nodal-line semimetals~\cite{Shao2018,Ahn2017}. However, neither of these two linear components extrapolate to zero conductivity at zero frequency, in conflict with the expected behavior for a 3D Dirac semi-metal. Generally, there are two reasons that linearly-increased optical conductivity does not extrapolate to the origin: (i)a gap in the massless Dirac band structure~\cite{Tabert2016} and (ii) the overlap of multiple interband transitions in Dirac like materials\cite{Shao2018}. The change of Fermi energy just causes a shift of the step at 2$|\mu|$, which has no effect on the extrapolation of the linear conductivity~\cite{Hosur2012,Ashby2014}. Multiple absorption peaks can be clearly observed at about 2\,000, 5\,000 and 9\,000~\icm\ in the optical conductivity, as indicated by arrows 1, 2 and 3, respectively. These peaks usually originate from inter-band transitions. The small Drude response and the limited value of the sharp absorption edge indicate that the Fermi level is not lying deeply in the valence/conduction bands.

In order to gain insights into the origin of these inter-band transitions, we have calculated the bulk band structure using first-principles methods. The crystal structure of \cms\ is depicted in Fig.~\ref{interband}(c). The band structure of \cms\ is calculated using density functional theory (DFT), employing the projector augmented wave (PAW) method encoded in the Vienna $ab$ initio simulation package(VASP)~\cite{Kresse1993, Kresse1996, Kresse1996-2}. The generalized-gradient approximation (GGA) for the exchange correlation function is used~\cite{John1996}. In the calculations we consider G-type antiferromagnetic, and the LDA+U method is used with the effective on-site Coulomb U being 5~eV for Mn 3$d$ states. Furthermore, considering the spin orbital coupling (SOC) in Sb atoms, SOC is included in our calculations. The experimental parameters ($a=22.09$\AA, $b=4.32$\AA, $c=4.35$\AA) were used in the calculations~\cite{He2017}.

Figure~\ref{interband}(b) shows the highly anisotropic band dispersion of \cms. The bands near the Fermi level are almost exclusively Sb derived, and the red lines are related to the density of states of Sb $5p$ orbitals. The inset of Fig.~\ref{interband}(b) shows the enlarged view of the Fermi level at the $Y$ point. Two different conduction bands cross the Fermi level, giving rise to two small electronic pockets. The presence of two electron Fermi surfaces suggests that our two-Drude model analysis captures the physics of this material. The small volume enclosed by the Fermi surface is in good agreement with the small Drude weight in our measured optical conductivity. Furthermore, we notice that all bands near the Fermi level are gapped and no Dirac points can be identified, demonstrating the absence of 3D Dirac fermions in the bulk states of \cms. This is in accordance with the conclusion we have reached from our experimental optical conductivity.

From the calculated band structure, we can easily resolve several interband transitions between different valence and conduction bands, such as 1-1$^{\prime}$, 1-2$^{\prime}$, 3-3$^{\prime}$, and 4-4$^{\prime}$. The onset energies of these transitions are 0.25, 0.35, 0.58, and 0.89~eV, respectively. Within the reasonable error, these energies agree quite well with the positions of the absorption peaks in the measured optical conductivity, implying that the optical conductivity in the high energy range is a result of the interband transitions. However, we notice that the transition at the $\Gamma$ point (0.29~eV) does not give rise to a noticeable peak feature in the optical spectrum, the reason is that these bands arising from the Sb $p_z$ orbitals have very low density of states.

At the $Y$ point, the valence bands (labeled with 1) and conduction bands (labeled with 1$^{\prime}$ and 2$^{\prime}$) disperse almost linearly in the momentum space. It is natural to assign the observed linear optical conductivity to the interband transitions between these bands. Further evidence may be revealed by a comparison of the onset energy of the linear optical conductivity and the band gaps. The gaps between these bands have the values of 0.25 and 0.35~eV, which agree well with the onset energies of the two linear components, strongly suggesting that the linear optical conductivity between 0.25 and 0.35~eV stems from the interband transitions between 1 and 1$^{\prime}$ bands, and the linear component between 0.35 and 0.6~eV is associated with the interband transitions from 1 to 1$^{\prime}$ and 2$^{\prime}$ bands.

\subsection{Band inversion}

\begin{figure*}[tb]
\centerline{
\includegraphics[width=2\columnwidth]{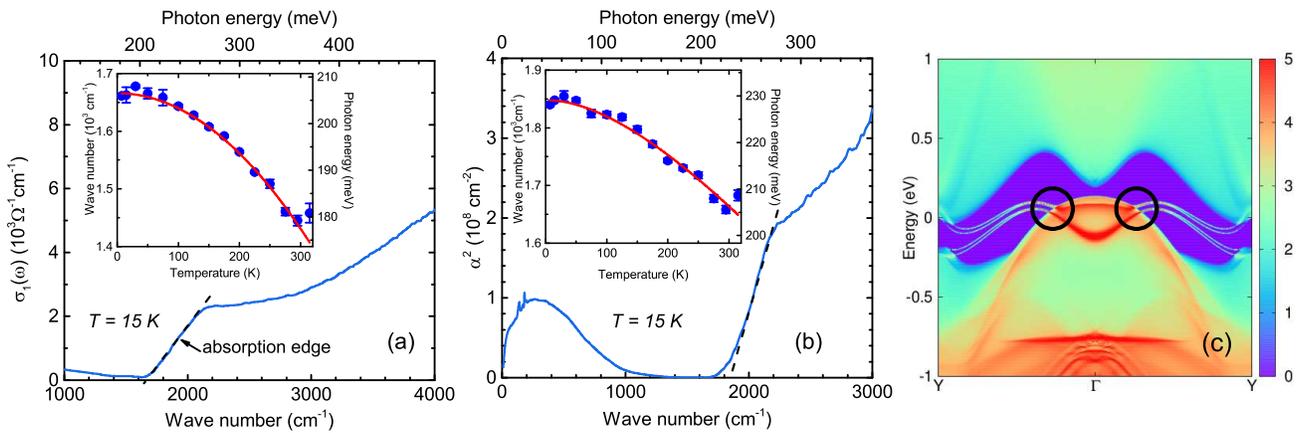}
}
\caption{(color online) (a) The linear extrapolation of absorption edge to estimate the gap energy. (b) The linear extrapolation of squared absorption coefficient to determine the gap energy. Both insets are the temperature dependence of the gap energy. The red solid lines denote the Varshni relation fit. (c) The electronic structure of surface state through Wannier function.}
\label{band inversion}
\end{figure*}

Although both our optical results and band structure calculations do not support the presence of 3D Dirac fermions in \cms, non-trivial properties have been reported by transport studies~\cite{He2017}. In order to better understand the properties of this material, further analysis of our optical data and band structure is necessary. A close examination of the calculated band structure [Fig.~\ref{interband}(b)] reveals a Mexican-hat-like dispersion of the electronic bands (labeled as 1$^{\prime}$) at the $Y$ point. This unusual feature in the band structure may arise from inverted bands~\cite{He2014}. It has been demonstrated that a Mexican-hat-like dispersion induced by the inverted bands can produce a van Hove singularity (VHS) in the optical conductivity spectrum (or joint density of states)~\cite{He2014,Xi2014}. Note that such a VHS feature can be clearly seen in our measured optical conductivity as shown by Arrow 1 in Fig.~\ref{interband}(a). Moreover, the optical conductivity of \cms\ exhibits exactly the same feature as the one calculated for inverted bands in Ref.~\cite{Xi2014}, supporting our speculation that band inversion may occur in the bands at the $Y$ point in this material.

Previous studies on Pb$_{1-x}$Sn$_{x}$Te, PbS$_{1-x}$Se$_{x}$, Pb$_{1-x}$Sn$_{x}$Se~\cite{Preier1979} have shown that for inverted bands, the band gap increases upon cooling as the lattice parameters are reduced with decreasing temperature. Such a band gap change can be easily detected by optical conductivity~\cite{Xi2014}. Following this line, further evidence for band inversion in \cms\ may be revealed by looking into the temperature dependence of the optical conductivity spectra. A sharp absorption edge can be seen between 1\,500 and 2\,000~\icm\ in the optical conductivity at all measured temperatures [see Fig.~\ref{band inversion}(a)]. At each temperature, the $x$-axis intercept of the extrapolation of the absorption edge yields the value of the band gap~\cite{Tokura2008}. The inset of Fig.~\ref{band inversion}(a) shows the temperature dependence of the band gap, which increases as the temperature is lowered, as expected for inverted bands.

We can also extract the evolution of the band gap via the squared absorption coefficient:
\begin{equation}
\label{absorption}
  \alpha^{2}=4\omega^{2}\kappa^{2}/c^{2},
\end{equation}
where $\kappa$ is the extinction coefficient which can be obtained from the Kramers-Kronig analysis of $R(\omega)$; $c$ is the speed of light. The band gap obtained from the linear extrapolation of $\alpha^{2}({\omega})$ to the $x$-axis for each temperature~\cite{Gu2014,Chen2017} is shown in Fig.~\ref{band inversion}(b). The gap increases upon cooling as depicted in the inset of Fig.~\ref{band inversion}(b). While the value of the gap obtained by this method is slightly larger than that obtained from the optical conductivity directly, they exhibit identical temperature dependence. The evolution of the band gap as a function of temperature follows the Varshni empirical relation(red dashed line through the data points)~\cite{Varshni1967}:
\begin{equation}
\label{varshni}
  E(T)=E(0)-\frac{\alpha T^2}{T+\beta},
\end{equation}
where $E(0)$ is the gap value at 0~K,  $\alpha$ is a constant, and $\beta$ is a parameter related to the Debye temperature.

For 3D topological insulator, albeit the bulk states are those of a trivial insulator, the surface should have gapless non-trivial edge states\cite{Hasan2010}. In order to investigate the topological properties of \cms, the surface states can be calculated by the surface Green function of the semi-infinite system using an iterative method. Fig.~\ref{band inversion}(c) shows the edge states on the [100] surface. A small gap can be identified along $\Gamma \sim Y$, as pointed by the black solid circles, indicating that the surface states are topologically trivial.

How can topological trivial states of \cms\ show the nontrivial properties in the magneto-transport measurements~\cite{He2017}?  In order to understand this, we can make a comparison to the results obtained in three-dimensional weak topological insulator \cfa\ with similar crystal structure\cite{Wu2015}. As shown in Ref.~\cite{Wu2015}, the As atoms in CaAs layer form zigzag chains, and behave like a Z$_2$ topologically nontrivial two-dimensional quantum spin Hall(QSH) insulator. However, one unit cell of \cms\ has two different Sb zigzag chains, as depicted in Fig.~\ref{interband}(c) (Sb1 and Sb3). As analyzed in Ref.~\cite{Wu2015}, the two different CaSb layers are Z$_2$ topologically nontrivial two-dimensional QSH insulators, and can lead to two sets of surface states on the [100] surface.  Owing to the absence of symmetry protection, the coupling between two sets of surface states is not prohibited, and the magnitude depends on the distance between two different CaSb layers, indicating that \cms\ is topological trivial. If the coupling between two sets of surface states is ignored, the non-robust gapless surface states can be identified. Hence, a gap about 25~meV is observed along $\Gamma \sim Y$ in Fig.~\ref{band inversion}(c), and the gap is so small that non-trivial properties may be measured in experiments.

\section{Conclusion}
To summarize, we have measured the detailed temperature and frequency dependence of the optical conductivity for \cms. The low-frequency optical conductivity is well described by the superposition of a narrow and a broad Drude components, which implies the existence of multiple Fermi surfaces. We observed several segments in the low-temperature optical conductivity varying linearly with frequency, but none of them extrapolates to the origin, which agrees well with the gapped nature of the electronic band structure obtained from first-principles calculations. A series of absorption peaks have been resolved in the high-frequency optical conductivity. The energies of these peaks are consistent with the calculated interband transitions. Both the line shape of the 2000-\icm\ absorption peak in our measured optical conductivity and band structure calculations point to Van Hove singularity behavior, suggesting that the bands at the $Y$ point are likely inverted. In addition, the smallest band gap increases upon cooling, mimic the temperature evolution of inverted bands as widely observed in topological crystalline insulators. Finally, our theoretical calculations demonstrate the existence of non-robust gapless surface states when the coupling between two Sb-chains layers is ignored, and the gap is so small that the non-trivial properties can be measured, even though \cms\ is topological trivial.

\section{Acknowledgments}
We thank Hongming Weng, Shunqing Shen and Kai Wang for useful discussions. This work was supported by NSFC (Projects No. 11774400 and No. 11404175) and MOST(973 Projects No. 2015CB921303, No. 2017YFA0302903 and No. 2015CB921102).


%
\bibliographystyle{apsrev}
\bibliography{biblio}

\begin{thebibliography}{54}
\expandafter\ifx\csname natexlab\endcsname\relax\def\natexlab#1{#1}\fi
\expandafter\ifx\csname bibnamefont\endcsname\relax
  \def\bibnamefont#1{#1}\fi
\expandafter\ifx\csname bibfnamefont\endcsname\relax
  \def\bibfnamefont#1{#1}\fi
\expandafter\ifx\csname citenamefont\endcsname\relax
  \def\citenamefont#1{#1}\fi
\expandafter\ifx\csname url\endcsname\relax
  \def\url#1{\texttt{#1}}\fi
\expandafter\ifx\csname urlprefix\endcsname\relax\def\urlprefix{URL }\fi
\providecommand{\bibinfo}[2]{#2}
\providecommand{\eprint}[2][]{\url{#2}}

\bibitem[{\citenamefont{Xi et~al.}(2014)\citenamefont{Xi, He, Guan, Liu, Zhong,
  Schneeloch, Liu, Gu, Du, Chen et~al.}}]{Xi2014}
\bibinfo{author}{\bibfnamefont{X.}~\bibnamefont{Xi}},
  \bibinfo{author}{\bibfnamefont{X.-G.} \bibnamefont{He}},
  \bibinfo{author}{\bibfnamefont{F.}~\bibnamefont{Guan}},
  \bibinfo{author}{\bibfnamefont{Z.}~\bibnamefont{Liu}},
  \bibinfo{author}{\bibfnamefont{R.~D.} \bibnamefont{Zhong}},
  \bibinfo{author}{\bibfnamefont{J.~A.} \bibnamefont{Schneeloch}},
  \bibinfo{author}{\bibfnamefont{T.~S.} \bibnamefont{Liu}},
  \bibinfo{author}{\bibfnamefont{G.~D.} \bibnamefont{Gu}},
  \bibinfo{author}{\bibfnamefont{X.}~\bibnamefont{Du}},
  \bibinfo{author}{\bibfnamefont{Z.}~\bibnamefont{Chen}}, \bibnamefont{et~al.},
  \bibinfo{journal}{Phys. Rev. Lett.} \textbf{\bibinfo{volume}{113}},
  \bibinfo{pages}{096401} (\bibinfo{year}{2014}).

\bibitem[{\citenamefont{Dziawa et~al.}(2012)\citenamefont{Dziawa, Kowalski,
  Dybko, Buczko, Szczerbakow, Szot, {\L}usakowska, Balasubramanian, Wojek,
  Berntsen et~al.}}]{Dziawa2012}
\bibinfo{author}{\bibfnamefont{P.}~\bibnamefont{Dziawa}},
  \bibinfo{author}{\bibfnamefont{B.~J.} \bibnamefont{Kowalski}},
  \bibinfo{author}{\bibfnamefont{K.}~\bibnamefont{Dybko}},
  \bibinfo{author}{\bibfnamefont{R.}~\bibnamefont{Buczko}},
  \bibinfo{author}{\bibfnamefont{A.}~\bibnamefont{Szczerbakow}},
  \bibinfo{author}{\bibfnamefont{M.}~\bibnamefont{Szot}},
  \bibinfo{author}{\bibfnamefont{E.}~\bibnamefont{{\L}usakowska}},
  \bibinfo{author}{\bibfnamefont{T.}~\bibnamefont{Balasubramanian}},
  \bibinfo{author}{\bibfnamefont{B.~M.} \bibnamefont{Wojek}},
  \bibinfo{author}{\bibfnamefont{M.~H.} \bibnamefont{Berntsen}},
  \bibnamefont{et~al.}, \bibinfo{journal}{Nat. Mater.}
  \textbf{\bibinfo{volume}{11}}, \bibinfo{pages}{1023} (\bibinfo{year}{2012}).

\bibitem[{\citenamefont{Zhang et~al.}(2009)\citenamefont{Zhang, Liu, Qi, Dai,
  Fang, and Zhang}}]{Zhang2009}
\bibinfo{author}{\bibfnamefont{H.}~\bibnamefont{Zhang}},
  \bibinfo{author}{\bibfnamefont{C.-X.} \bibnamefont{Liu}},
  \bibinfo{author}{\bibfnamefont{X.-L.} \bibnamefont{Qi}},
  \bibinfo{author}{\bibfnamefont{X.}~\bibnamefont{Dai}},
  \bibinfo{author}{\bibfnamefont{Z.}~\bibnamefont{Fang}}, \bibnamefont{and}
  \bibinfo{author}{\bibfnamefont{S.-C.} \bibnamefont{Zhang}},
  \bibinfo{journal}{Nat. Phys.} \textbf{\bibinfo{volume}{5}},
  \bibinfo{pages}{438} (\bibinfo{year}{2009}).

\bibitem[{\citenamefont{Wang et~al.}(2012{\natexlab{a}})\citenamefont{Wang,
  Sun, Chen, Franchini, Xu, Weng, Dai, and Fang}}]{wang2012}
\bibinfo{author}{\bibfnamefont{Z.}~\bibnamefont{Wang}},
  \bibinfo{author}{\bibfnamefont{Y.}~\bibnamefont{Sun}},
  \bibinfo{author}{\bibfnamefont{X.-Q.} \bibnamefont{Chen}},
  \bibinfo{author}{\bibfnamefont{C.}~\bibnamefont{Franchini}},
  \bibinfo{author}{\bibfnamefont{G.}~\bibnamefont{Xu}},
  \bibinfo{author}{\bibfnamefont{H.}~\bibnamefont{Weng}},
  \bibinfo{author}{\bibfnamefont{X.}~\bibnamefont{Dai}}, \bibnamefont{and}
  \bibinfo{author}{\bibfnamefont{Z.}~\bibnamefont{Fang}},
  \bibinfo{journal}{Phys. Rev. B} \textbf{\bibinfo{volume}{85}},
  \bibinfo{pages}{195320} (\bibinfo{year}{2012}{\natexlab{a}}).

\bibitem[{\citenamefont{Wang et~al.}(2013)\citenamefont{Wang, Weng, Wu, Dai,
  and Fang}}]{wang2013}
\bibinfo{author}{\bibfnamefont{Z.}~\bibnamefont{Wang}},
  \bibinfo{author}{\bibfnamefont{H.}~\bibnamefont{Weng}},
  \bibinfo{author}{\bibfnamefont{Q.}~\bibnamefont{Wu}},
  \bibinfo{author}{\bibfnamefont{X.}~\bibnamefont{Dai}}, \bibnamefont{and}
  \bibinfo{author}{\bibfnamefont{Z.}~\bibnamefont{Fang}},
  \bibinfo{journal}{Phys. Rev. B} \textbf{\bibinfo{volume}{88}},
  \bibinfo{pages}{125427} (\bibinfo{year}{2013}).

\bibitem[{\citenamefont{Akrap et~al.}(2016)\citenamefont{Akrap, Hakl,
  Tchoumakov, Crassee, Kuba, Goerbig, Homes, Caha, Nov\'ak, Teppe
  et~al.}}]{Akrap2016}
\bibinfo{author}{\bibfnamefont{A.}~\bibnamefont{Akrap}},
  \bibinfo{author}{\bibfnamefont{M.}~\bibnamefont{Hakl}},
  \bibinfo{author}{\bibfnamefont{S.}~\bibnamefont{Tchoumakov}},
  \bibinfo{author}{\bibfnamefont{I.}~\bibnamefont{Crassee}},
  \bibinfo{author}{\bibfnamefont{J.}~\bibnamefont{Kuba}},
  \bibinfo{author}{\bibfnamefont{M.~O.} \bibnamefont{Goerbig}},
  \bibinfo{author}{\bibfnamefont{C.~C.} \bibnamefont{Homes}},
  \bibinfo{author}{\bibfnamefont{O.}~\bibnamefont{Caha}},
  \bibinfo{author}{\bibfnamefont{J.}~\bibnamefont{Nov\'ak}},
  \bibinfo{author}{\bibfnamefont{F.}~\bibnamefont{Teppe}},
  \bibnamefont{et~al.}, \bibinfo{journal}{Phys. Rev. Lett.}
  \textbf{\bibinfo{volume}{117}}, \bibinfo{pages}{136401}
  (\bibinfo{year}{2016}).

\bibitem[{\citenamefont{Weng et~al.}(2015)\citenamefont{Weng, Fang, Fang,
  Bernevig, and Dai}}]{weng2015}
\bibinfo{author}{\bibfnamefont{H.}~\bibnamefont{Weng}},
  \bibinfo{author}{\bibfnamefont{C.}~\bibnamefont{Fang}},
  \bibinfo{author}{\bibfnamefont{Z.}~\bibnamefont{Fang}},
  \bibinfo{author}{\bibfnamefont{B.~A.} \bibnamefont{Bernevig}},
  \bibnamefont{and} \bibinfo{author}{\bibfnamefont{X.}~\bibnamefont{Dai}},
  \bibinfo{journal}{Phys. Rev. X} \textbf{\bibinfo{volume}{5}},
  \bibinfo{pages}{011029} (\bibinfo{year}{2015}).

\bibitem[{\citenamefont{Huang et~al.}(2015)\citenamefont{Huang, Xu, Belopolski,
  Lee, Chang, Wang, Alidoust, Bian, Neupane, Zhang et~al.}}]{Huang2015}
\bibinfo{author}{\bibfnamefont{S.-M.} \bibnamefont{Huang}},
  \bibinfo{author}{\bibfnamefont{S.-Y.} \bibnamefont{Xu}},
  \bibinfo{author}{\bibfnamefont{I.}~\bibnamefont{Belopolski}},
  \bibinfo{author}{\bibfnamefont{C.-C.} \bibnamefont{Lee}},
  \bibinfo{author}{\bibfnamefont{G.}~\bibnamefont{Chang}},
  \bibinfo{author}{\bibfnamefont{B.}~\bibnamefont{Wang}},
  \bibinfo{author}{\bibfnamefont{N.}~\bibnamefont{Alidoust}},
  \bibinfo{author}{\bibfnamefont{G.}~\bibnamefont{Bian}},
  \bibinfo{author}{\bibfnamefont{M.}~\bibnamefont{Neupane}},
  \bibinfo{author}{\bibfnamefont{C.}~\bibnamefont{Zhang}},
  \bibnamefont{et~al.}, \bibinfo{journal}{Nat. Commun.}
  \textbf{\bibinfo{volume}{6}}, \bibinfo{pages}{7373} (\bibinfo{year}{2015}).

\bibitem[{\citenamefont{Bansil et~al.}(2016)\citenamefont{Bansil, Lin, and
  Das}}]{Bansil2016}
\bibinfo{author}{\bibfnamefont{A.}~\bibnamefont{Bansil}},
  \bibinfo{author}{\bibfnamefont{H.}~\bibnamefont{Lin}}, \bibnamefont{and}
  \bibinfo{author}{\bibfnamefont{T.}~\bibnamefont{Das}}, \bibinfo{journal}{Rev.
  Mod. Phys.} \textbf{\bibinfo{volume}{88}}, \bibinfo{pages}{021004}
  (\bibinfo{year}{2016}).

\bibitem[{\citenamefont{Hasan and Kane}(2010)}]{Hasan2010}
\bibinfo{author}{\bibfnamefont{M.~Z.} \bibnamefont{Hasan}} \bibnamefont{and}
  \bibinfo{author}{\bibfnamefont{C.~L.} \bibnamefont{Kane}},
  \bibinfo{journal}{Rev. Mod. Phys.} \textbf{\bibinfo{volume}{82}},
  \bibinfo{pages}{3045} (\bibinfo{year}{2010}).

\bibitem[{\citenamefont{Qi and Zhang}(2011)}]{Qi2011}
\bibinfo{author}{\bibfnamefont{X.-L.} \bibnamefont{Qi}} \bibnamefont{and}
  \bibinfo{author}{\bibfnamefont{S.-C.} \bibnamefont{Zhang}},
  \bibinfo{journal}{Rev. Mod. Phys.} \textbf{\bibinfo{volume}{83}},
  \bibinfo{pages}{1057} (\bibinfo{year}{2011}).

\bibitem[{\citenamefont{Park et~al.}(2011)\citenamefont{Park, Lee,
  Wolff-Fabris, Koh, Eom, Kim, Farhan, Jo, Kim, Shim et~al.}}]{Park2011}
\bibinfo{author}{\bibfnamefont{J.}~\bibnamefont{Park}},
  \bibinfo{author}{\bibfnamefont{G.}~\bibnamefont{Lee}},
  \bibinfo{author}{\bibfnamefont{F.}~\bibnamefont{Wolff-Fabris}},
  \bibinfo{author}{\bibfnamefont{Y.~Y.} \bibnamefont{Koh}},
  \bibinfo{author}{\bibfnamefont{M.~J.} \bibnamefont{Eom}},
  \bibinfo{author}{\bibfnamefont{Y.~K.} \bibnamefont{Kim}},
  \bibinfo{author}{\bibfnamefont{M.~A.} \bibnamefont{Farhan}},
  \bibinfo{author}{\bibfnamefont{Y.~J.} \bibnamefont{Jo}},
  \bibinfo{author}{\bibfnamefont{C.}~\bibnamefont{Kim}},
  \bibinfo{author}{\bibfnamefont{J.~H.} \bibnamefont{Shim}},
  \bibnamefont{et~al.}, \bibinfo{journal}{Phys. Rev. Lett.}
  \textbf{\bibinfo{volume}{107}}, \bibinfo{pages}{126402}
  (\bibinfo{year}{2011}).

\bibitem[{\citenamefont{Wang et~al.}(2012{\natexlab{b}})\citenamefont{Wang,
  Graf, Wang, Lei, Tozer, and Petrovic}}]{wangcmb2012}
\bibinfo{author}{\bibfnamefont{K.}~\bibnamefont{Wang}},
  \bibinfo{author}{\bibfnamefont{D.}~\bibnamefont{Graf}},
  \bibinfo{author}{\bibfnamefont{L.}~\bibnamefont{Wang}},
  \bibinfo{author}{\bibfnamefont{H.}~\bibnamefont{Lei}},
  \bibinfo{author}{\bibfnamefont{S.~W.} \bibnamefont{Tozer}}, \bibnamefont{and}
  \bibinfo{author}{\bibfnamefont{C.}~\bibnamefont{Petrovic}},
  \bibinfo{journal}{Phys. Rev. B} \textbf{\bibinfo{volume}{85}},
  \bibinfo{pages}{041101} (\bibinfo{year}{2012}{\natexlab{b}}).

\bibitem[{\citenamefont{Li et~al.}(2016)\citenamefont{Li, Wang, Graf, Wang,
  Wang, and Petrovic}}]{Li2016}
\bibinfo{author}{\bibfnamefont{L.}~\bibnamefont{Li}},
  \bibinfo{author}{\bibfnamefont{K.}~\bibnamefont{Wang}},
  \bibinfo{author}{\bibfnamefont{D.}~\bibnamefont{Graf}},
  \bibinfo{author}{\bibfnamefont{L.}~\bibnamefont{Wang}},
  \bibinfo{author}{\bibfnamefont{A.}~\bibnamefont{Wang}}, \bibnamefont{and}
  \bibinfo{author}{\bibfnamefont{C.}~\bibnamefont{Petrovic}},
  \bibinfo{journal}{Phys. Rev. B} \textbf{\bibinfo{volume}{93}},
  \bibinfo{pages}{115141} (\bibinfo{year}{2016}).

\bibitem[{\citenamefont{Wang et~al.}(2016)\citenamefont{Wang, Zaliznyak, Ren,
  Wu, Graf, Garlea, Warren, Bozin, Zhu, and Petrovic}}]{wangymb2016}
\bibinfo{author}{\bibfnamefont{A.}~\bibnamefont{Wang}},
  \bibinfo{author}{\bibfnamefont{I.}~\bibnamefont{Zaliznyak}},
  \bibinfo{author}{\bibfnamefont{W.}~\bibnamefont{Ren}},
  \bibinfo{author}{\bibfnamefont{L.}~\bibnamefont{Wu}},
  \bibinfo{author}{\bibfnamefont{D.}~\bibnamefont{Graf}},
  \bibinfo{author}{\bibfnamefont{V.~O.} \bibnamefont{Garlea}},
  \bibinfo{author}{\bibfnamefont{J.~B.} \bibnamefont{Warren}},
  \bibinfo{author}{\bibfnamefont{E.}~\bibnamefont{Bozin}},
  \bibinfo{author}{\bibfnamefont{Y.}~\bibnamefont{Zhu}}, \bibnamefont{and}
  \bibinfo{author}{\bibfnamefont{C.}~\bibnamefont{Petrovic}},
  \bibinfo{journal}{Phys. Rev. B} \textbf{\bibinfo{volume}{94}},
  \bibinfo{pages}{165161} (\bibinfo{year}{2016}).

\bibitem[{\citenamefont{Wang et~al.}(2011)\citenamefont{Wang, Graf, Lei, Tozer,
  and Petrovic}}]{wangsmb2011}
\bibinfo{author}{\bibfnamefont{K.}~\bibnamefont{Wang}},
  \bibinfo{author}{\bibfnamefont{D.}~\bibnamefont{Graf}},
  \bibinfo{author}{\bibfnamefont{H.}~\bibnamefont{Lei}},
  \bibinfo{author}{\bibfnamefont{S.~W.} \bibnamefont{Tozer}}, \bibnamefont{and}
  \bibinfo{author}{\bibfnamefont{C.}~\bibnamefont{Petrovic}},
  \bibinfo{journal}{Phys. Rev. B} \textbf{\bibinfo{volume}{84}},
  \bibinfo{pages}{220401} (\bibinfo{year}{2011}).

\bibitem[{\citenamefont{May et~al.}(2014)\citenamefont{May, McGuire, and
  Sales}}]{Andrew2014}
\bibinfo{author}{\bibfnamefont{A.~F.} \bibnamefont{May}},
  \bibinfo{author}{\bibfnamefont{M.~A.} \bibnamefont{McGuire}},
  \bibnamefont{and} \bibinfo{author}{\bibfnamefont{B.~C.} \bibnamefont{Sales}},
  \bibinfo{journal}{Phys. Rev. B} \textbf{\bibinfo{volume}{90}},
  \bibinfo{pages}{075109} (\bibinfo{year}{2014}).

\bibitem[{\citenamefont{Chinotti et~al.}(2016)\citenamefont{Chinotti, Pal, Ren,
  Petrovic, and Degiorgi}}]{Chinotti2016}
\bibinfo{author}{\bibfnamefont{M.}~\bibnamefont{Chinotti}},
  \bibinfo{author}{\bibfnamefont{A.}~\bibnamefont{Pal}},
  \bibinfo{author}{\bibfnamefont{W.~J.} \bibnamefont{Ren}},
  \bibinfo{author}{\bibfnamefont{C.}~\bibnamefont{Petrovic}}, \bibnamefont{and}
  \bibinfo{author}{\bibfnamefont{L.}~\bibnamefont{Degiorgi}},
  \bibinfo{journal}{Phys. Rev. B} \textbf{\bibinfo{volume}{94}},
  \bibinfo{pages}{245101} (\bibinfo{year}{2016}).

\bibitem[{\citenamefont{Lee et~al.}(2013)\citenamefont{Lee, Farhan, Kim, and
  Shim}}]{Lee2013}
\bibinfo{author}{\bibfnamefont{G.}~\bibnamefont{Lee}},
  \bibinfo{author}{\bibfnamefont{M.~A.} \bibnamefont{Farhan}},
  \bibinfo{author}{\bibfnamefont{J.~S.} \bibnamefont{Kim}}, \bibnamefont{and}
  \bibinfo{author}{\bibfnamefont{J.~H.} \bibnamefont{Shim}},
  \bibinfo{journal}{Phys. Rev. B} \textbf{\bibinfo{volume}{87}},
  \bibinfo{pages}{245104} (\bibinfo{year}{2013}).

\bibitem[{\citenamefont{Liu et~al.}(2016)\citenamefont{Liu, Hu, Cao, Zhu,
  Chuang, Graf, Adams, Radmanesh, Spinu, Chiorescu et~al.}}]{Liu2016}
\bibinfo{author}{\bibfnamefont{J.}~\bibnamefont{Liu}},
  \bibinfo{author}{\bibfnamefont{J.}~\bibnamefont{Hu}},
  \bibinfo{author}{\bibfnamefont{H.}~\bibnamefont{Cao}},
  \bibinfo{author}{\bibfnamefont{Y.}~\bibnamefont{Zhu}},
  \bibinfo{author}{\bibfnamefont{A.}~\bibnamefont{Chuang}},
  \bibinfo{author}{\bibfnamefont{D.}~\bibnamefont{Graf}},
  \bibinfo{author}{\bibfnamefont{D.~J.} \bibnamefont{Adams}},
  \bibinfo{author}{\bibfnamefont{S.~M.~A.} \bibnamefont{Radmanesh}},
  \bibinfo{author}{\bibfnamefont{L.}~\bibnamefont{Spinu}},
  \bibinfo{author}{\bibfnamefont{I.}~\bibnamefont{Chiorescu}},
  \bibnamefont{et~al.}, \bibinfo{journal}{Sci. Rep.}
  \textbf{\bibinfo{volume}{6}}, \bibinfo{pages}{30525} (\bibinfo{year}{2016}).

\bibitem[{\citenamefont{Chaudhuri et~al.}(2017)\citenamefont{Chaudhuri, Cheng,
  Yaresko, Gibson, Cava, and Armitage}}]{Chaudhuri2017}
\bibinfo{author}{\bibfnamefont{D.}~\bibnamefont{Chaudhuri}},
  \bibinfo{author}{\bibfnamefont{B.}~\bibnamefont{Cheng}},
  \bibinfo{author}{\bibfnamefont{A.}~\bibnamefont{Yaresko}},
  \bibinfo{author}{\bibfnamefont{Q.~D.} \bibnamefont{Gibson}},
  \bibinfo{author}{\bibfnamefont{R.~J.} \bibnamefont{Cava}}, \bibnamefont{and}
  \bibinfo{author}{\bibfnamefont{N.~P.} \bibnamefont{Armitage}},
  \bibinfo{journal}{Phys. Rev. B} \textbf{\bibinfo{volume}{96}},
  \bibinfo{pages}{075151} (\bibinfo{year}{2017}).

\bibitem[{\citenamefont{He et~al.}(2012)\citenamefont{He, Wang, and
  Chen}}]{He2012}
\bibinfo{author}{\bibfnamefont{J.~B.} \bibnamefont{He}},
  \bibinfo{author}{\bibfnamefont{D.~M.} \bibnamefont{Wang}}, \bibnamefont{and}
  \bibinfo{author}{\bibfnamefont{G.~F.} \bibnamefont{Chen}},
  \bibinfo{journal}{App. Phys. Lett.} \textbf{\bibinfo{volume}{100}},
  \bibinfo{pages}{112405} (\bibinfo{year}{2012}).

\bibitem[{\citenamefont{Borisenko et~al.}(2015)\citenamefont{Borisenko,
  Evtushinsky, Gibson, Yaresko, Kim, Ali, Buechner, Hoesch, and
  Cava}}]{Borisenko2015}
\bibinfo{author}{\bibfnamefont{S.}~\bibnamefont{Borisenko}},
  \bibinfo{author}{\bibfnamefont{D.}~\bibnamefont{Evtushinsky}},
  \bibinfo{author}{\bibfnamefont{Q.}~\bibnamefont{Gibson}},
  \bibinfo{author}{\bibfnamefont{A.}~\bibnamefont{Yaresko}},
  \bibinfo{author}{\bibfnamefont{T.}~\bibnamefont{Kim}},
  \bibinfo{author}{\bibfnamefont{M.~N.} \bibnamefont{Ali}},
  \bibinfo{author}{\bibfnamefont{B.}~\bibnamefont{Buechner}},
  \bibinfo{author}{\bibfnamefont{M.}~\bibnamefont{Hoesch}}, \bibnamefont{and}
  \bibinfo{author}{\bibfnamefont{R.~J.} \bibnamefont{Cava}}
  (\bibinfo{year}{2015}), \eprint{arXiv: 1507.04847}.

\bibitem[{\citenamefont{Farhan et~al.}(2014)\citenamefont{Farhan, Lee, and
  Shim}}]{Farhan2014}
\bibinfo{author}{\bibfnamefont{M.~A.} \bibnamefont{Farhan}},
  \bibinfo{author}{\bibfnamefont{G.}~\bibnamefont{Lee}}, \bibnamefont{and}
  \bibinfo{author}{\bibfnamefont{J.~H.} \bibnamefont{Shim}},
  \bibinfo{journal}{J. Phys.: Condens. Matter} \textbf{\bibinfo{volume}{26}},
  \bibinfo{pages}{042201} (\bibinfo{year}{2014}).

\bibitem[{\citenamefont{Liu et~al.}(2015)\citenamefont{Liu, Hu, Zhang, Graf,
  Cao, Radmanesh, Adams, Zhu, Cheng, Liu et~al.}}]{Liu2015}
\bibinfo{author}{\bibfnamefont{J.~Y.} \bibnamefont{Liu}},
  \bibinfo{author}{\bibfnamefont{J.}~\bibnamefont{Hu}},
  \bibinfo{author}{\bibfnamefont{Q.}~\bibnamefont{Zhang}},
  \bibinfo{author}{\bibfnamefont{D.}~\bibnamefont{Graf}},
  \bibinfo{author}{\bibfnamefont{H.~B.} \bibnamefont{Cao}},
  \bibinfo{author}{\bibfnamefont{S.~M.~A.} \bibnamefont{Radmanesh}},
  \bibinfo{author}{\bibfnamefont{D.~J.} \bibnamefont{Adams}},
  \bibinfo{author}{\bibfnamefont{Y.~L.} \bibnamefont{Zhu}},
  \bibinfo{author}{\bibfnamefont{G.~F.} \bibnamefont{Cheng}},
  \bibinfo{author}{\bibfnamefont{X.}~\bibnamefont{Liu}}, \bibnamefont{et~al.}
  (\bibinfo{year}{2015}), \eprint{arXiv: 1507.07978}.

\bibitem[{\citenamefont{Brechtel et~al.}(1981)\citenamefont{Brechtel, Cordier,
  and Schafer}}]{BRECHTEL1981131}
\bibinfo{author}{\bibfnamefont{E.}~\bibnamefont{Brechtel}},
  \bibinfo{author}{\bibfnamefont{G.}~\bibnamefont{Cordier}}, \bibnamefont{and}
  \bibinfo{author}{\bibfnamefont{H.}~\bibnamefont{Schafer}},
  \bibinfo{journal}{J. Less-Common Met.} \textbf{\bibinfo{volume}{79,131}}
  (\bibinfo{year}{1981}).

\bibitem[{\citenamefont{Ramankutty et~al.}(2017)\citenamefont{Ramankutty,
  Henke, Schiphorst, Nutakki, Bron, Araizi-Kanoutas, Mishra, Li, Huang, Kim
  et~al.}}]{Ramankutty2017}
\bibinfo{author}{\bibfnamefont{S.~V.} \bibnamefont{Ramankutty}},
  \bibinfo{author}{\bibfnamefont{J.}~\bibnamefont{Henke}},
  \bibinfo{author}{\bibfnamefont{A.}~\bibnamefont{Schiphorst}},
  \bibinfo{author}{\bibfnamefont{R.}~\bibnamefont{Nutakki}},
  \bibinfo{author}{\bibfnamefont{S.}~\bibnamefont{Bron}},
  \bibinfo{author}{\bibfnamefont{G.}~\bibnamefont{Araizi-Kanoutas}},
  \bibinfo{author}{\bibfnamefont{S.~K.} \bibnamefont{Mishra}},
  \bibinfo{author}{\bibfnamefont{L.}~\bibnamefont{Li}},
  \bibinfo{author}{\bibfnamefont{Y.~K.} \bibnamefont{Huang}},
  \bibinfo{author}{\bibfnamefont{T.~K.} \bibnamefont{Kim}},
  \bibnamefont{et~al.} (\bibinfo{year}{2017}), \eprint{arXiv: 1711.07165}.

\bibitem[{\citenamefont{He et~al.}(2017)\citenamefont{He, Fu, Zhao, Liang,
  Chen, Leng, Wang, Li, Zhang, Xue et~al.}}]{He2017}
\bibinfo{author}{\bibfnamefont{J.~B.} \bibnamefont{He}},
  \bibinfo{author}{\bibfnamefont{Y.}~\bibnamefont{Fu}},
  \bibinfo{author}{\bibfnamefont{L.~X.} \bibnamefont{Zhao}},
  \bibinfo{author}{\bibfnamefont{H.}~\bibnamefont{Liang}},
  \bibinfo{author}{\bibfnamefont{D.}~\bibnamefont{Chen}},
  \bibinfo{author}{\bibfnamefont{Y.~M.} \bibnamefont{Leng}},
  \bibinfo{author}{\bibfnamefont{X.~M.} \bibnamefont{Wang}},
  \bibinfo{author}{\bibfnamefont{J.}~\bibnamefont{Li}},
  \bibinfo{author}{\bibfnamefont{S.}~\bibnamefont{Zhang}},
  \bibinfo{author}{\bibfnamefont{M.~Q.} \bibnamefont{Xue}},
  \bibnamefont{et~al.}, \bibinfo{journal}{Phys. Rev. B}
  \textbf{\bibinfo{volume}{95}}, \bibinfo{pages}{045128}
  (\bibinfo{year}{2017}).

\bibitem[{\citenamefont{Hosur et~al.}(2012)\citenamefont{Hosur, Parameswaran,
  and Vishwanath}}]{Hosur2012}
\bibinfo{author}{\bibfnamefont{P.}~\bibnamefont{Hosur}},
  \bibinfo{author}{\bibfnamefont{S.~A.} \bibnamefont{Parameswaran}},
  \bibnamefont{and}
  \bibinfo{author}{\bibfnamefont{A.}~\bibnamefont{Vishwanath}},
  \bibinfo{journal}{Phys. Rev. Lett.} \textbf{\bibinfo{volume}{108}},
  \bibinfo{pages}{046602} (\bibinfo{year}{2012}).

\bibitem[{\citenamefont{Ashby and Carbotte}(2014)}]{Ashby2014}
\bibinfo{author}{\bibfnamefont{P.~E.~C.} \bibnamefont{Ashby}} \bibnamefont{and}
  \bibinfo{author}{\bibfnamefont{J.~P.} \bibnamefont{Carbotte}},
  \bibinfo{journal}{Phys. Rev. B} \textbf{\bibinfo{volume}{89}},
  \bibinfo{pages}{245121} (\bibinfo{year}{2014}).

\bibitem[{\citenamefont{Armitage et~al.}(2018)\citenamefont{Armitage, Mele, and
  Vishwanath}}]{Aemitage2018}
\bibinfo{author}{\bibfnamefont{N.~P.} \bibnamefont{Armitage}},
  \bibinfo{author}{\bibfnamefont{E.~J.} \bibnamefont{Mele}}, \bibnamefont{and}
  \bibinfo{author}{\bibfnamefont{A.}~\bibnamefont{Vishwanath}},
  \bibinfo{journal}{Rev. Mod. Phys.} \textbf{\bibinfo{volume}{90}},
  \bibinfo{pages}{015001} (\bibinfo{year}{2018}).

\bibitem[{\citenamefont{Homes et~al.}(1993)\citenamefont{Homes, Reedyk,
  Cradles, and Timusk}}]{Homes1993}
\bibinfo{author}{\bibfnamefont{C.~C.} \bibnamefont{Homes}},
  \bibinfo{author}{\bibfnamefont{M.}~\bibnamefont{Reedyk}},
  \bibinfo{author}{\bibfnamefont{D.~A.} \bibnamefont{Cradles}},
  \bibnamefont{and} \bibinfo{author}{\bibfnamefont{T.}~\bibnamefont{Timusk}},
  \bibinfo{journal}{Appl. Opt.} \textbf{\bibinfo{volume}{32}}
  (\bibinfo{year}{1993}).

\bibitem[{\citenamefont{Schilling
  et~al.}(2017{\natexlab{a}})\citenamefont{Schilling, L\"ohle, Neubauer,
  Shekhar, Felser, Dressel, and Pronin}}]{Schilling2017}
\bibinfo{author}{\bibfnamefont{M.~B.} \bibnamefont{Schilling}},
  \bibinfo{author}{\bibfnamefont{A.}~\bibnamefont{L\"ohle}},
  \bibinfo{author}{\bibfnamefont{D.}~\bibnamefont{Neubauer}},
  \bibinfo{author}{\bibfnamefont{C.}~\bibnamefont{Shekhar}},
  \bibinfo{author}{\bibfnamefont{C.}~\bibnamefont{Felser}},
  \bibinfo{author}{\bibfnamefont{M.}~\bibnamefont{Dressel}}, \bibnamefont{and}
  \bibinfo{author}{\bibfnamefont{A.~V.} \bibnamefont{Pronin}},
  \bibinfo{journal}{Phys. Rev. B} \textbf{\bibinfo{volume}{95}},
  \bibinfo{pages}{155201} (\bibinfo{year}{2017}{\natexlab{a}}).

\bibitem[{\citenamefont{Beach and Christy}(1977)}]{Beach1977}
\bibinfo{author}{\bibfnamefont{R.~T.} \bibnamefont{Beach}} \bibnamefont{and}
  \bibinfo{author}{\bibfnamefont{R.~W.} \bibnamefont{Christy}},
  \bibinfo{journal}{Phys. Rev. B} \textbf{\bibinfo{volume}{16}},
  \bibinfo{pages}{5277} (\bibinfo{year}{1977}).

\bibitem[{\citenamefont{B\'acsi and Virosztek}(2013)}]{Bacsi2013}
\bibinfo{author}{\bibfnamefont{A.}~\bibnamefont{B\'acsi}} \bibnamefont{and}
  \bibinfo{author}{\bibfnamefont{A.}~\bibnamefont{Virosztek}},
  \bibinfo{journal}{Phys. Rev. B} \textbf{\bibinfo{volume}{87}},
  \bibinfo{pages}{125425} (\bibinfo{year}{2013}).

\bibitem[{\citenamefont{Schilling
  et~al.}(2017{\natexlab{b}})\citenamefont{Schilling, Schoop, Lotsch, Dressel,
  and Pronin}}]{Schilling2017-2}
\bibinfo{author}{\bibfnamefont{M.~B.} \bibnamefont{Schilling}},
  \bibinfo{author}{\bibfnamefont{L.~M.} \bibnamefont{Schoop}},
  \bibinfo{author}{\bibfnamefont{B.~V.} \bibnamefont{Lotsch}},
  \bibinfo{author}{\bibfnamefont{M.}~\bibnamefont{Dressel}}, \bibnamefont{and}
  \bibinfo{author}{\bibfnamefont{A.~V.} \bibnamefont{Pronin}},
  \bibinfo{journal}{Phys. Rev. Lett.} \textbf{\bibinfo{volume}{119}},
  \bibinfo{pages}{187401} (\bibinfo{year}{2017}{\natexlab{b}}).

\bibitem[{\citenamefont{Dai et~al.}(2014)\citenamefont{Dai, Akrap, Schneeloch,
  Zhong, Liu, Gu, Li, and Homes}}]{Dai2014}
\bibinfo{author}{\bibfnamefont{Y.~M.} \bibnamefont{Dai}},
  \bibinfo{author}{\bibfnamefont{A.}~\bibnamefont{Akrap}},
  \bibinfo{author}{\bibfnamefont{J.}~\bibnamefont{Schneeloch}},
  \bibinfo{author}{\bibfnamefont{R.~D.} \bibnamefont{Zhong}},
  \bibinfo{author}{\bibfnamefont{T.~S.} \bibnamefont{Liu}},
  \bibinfo{author}{\bibfnamefont{G.~D.} \bibnamefont{Gu}},
  \bibinfo{author}{\bibfnamefont{Q.}~\bibnamefont{Li}}, \bibnamefont{and}
  \bibinfo{author}{\bibfnamefont{C.~C.} \bibnamefont{Homes}},
  \bibinfo{journal}{Phys. Rev. B} \textbf{\bibinfo{volume}{90}},
  \bibinfo{pages}{121114} (\bibinfo{year}{2014}).

\bibitem[{\citenamefont{Neubauer et~al.}(2016)\citenamefont{Neubauer, Carbotte,
  Nateprov, L\"ohle, Dressel, and Pronin}}]{Neubauer2016}
\bibinfo{author}{\bibfnamefont{D.}~\bibnamefont{Neubauer}},
  \bibinfo{author}{\bibfnamefont{J.~P.} \bibnamefont{Carbotte}},
  \bibinfo{author}{\bibfnamefont{A.~A.} \bibnamefont{Nateprov}},
  \bibinfo{author}{\bibfnamefont{A.}~\bibnamefont{L\"ohle}},
  \bibinfo{author}{\bibfnamefont{M.}~\bibnamefont{Dressel}}, \bibnamefont{and}
  \bibinfo{author}{\bibfnamefont{A.~V.} \bibnamefont{Pronin}},
  \bibinfo{journal}{Phys. Rev. B} \textbf{\bibinfo{volume}{93}},
  \bibinfo{pages}{121202} (\bibinfo{year}{2016}).

\bibitem[{\citenamefont{Chen et~al.}(2015)\citenamefont{Chen, Zhang,
  Schneeloch, Zhang, Li, Gu, and Wang}}]{chen2015}
\bibinfo{author}{\bibfnamefont{R.~Y.} \bibnamefont{Chen}},
  \bibinfo{author}{\bibfnamefont{S.~J.} \bibnamefont{Zhang}},
  \bibinfo{author}{\bibfnamefont{J.~A.} \bibnamefont{Schneeloch}},
  \bibinfo{author}{\bibfnamefont{C.}~\bibnamefont{Zhang}},
  \bibinfo{author}{\bibfnamefont{Q.}~\bibnamefont{Li}},
  \bibinfo{author}{\bibfnamefont{G.~D.} \bibnamefont{Gu}}, \bibnamefont{and}
  \bibinfo{author}{\bibfnamefont{N.~L.} \bibnamefont{Wang}},
  \bibinfo{journal}{Phys. Rev. B} \textbf{\bibinfo{volume}{92}},
  \bibinfo{pages}{075107} (\bibinfo{year}{2015}).

\bibitem[{\citenamefont{Xu et~al.}(2016)\citenamefont{Xu, Dai, Zhao, Wang,
  Yang, Zhang, Liu, Xiao, Chen, Taylor et~al.}}]{Xu2016}
\bibinfo{author}{\bibfnamefont{B.}~\bibnamefont{Xu}},
  \bibinfo{author}{\bibfnamefont{Y.~M.} \bibnamefont{Dai}},
  \bibinfo{author}{\bibfnamefont{L.~X.} \bibnamefont{Zhao}},
  \bibinfo{author}{\bibfnamefont{K.}~\bibnamefont{Wang}},
  \bibinfo{author}{\bibfnamefont{R.}~\bibnamefont{Yang}},
  \bibinfo{author}{\bibfnamefont{W.}~\bibnamefont{Zhang}},
  \bibinfo{author}{\bibfnamefont{J.~Y.} \bibnamefont{Liu}},
  \bibinfo{author}{\bibfnamefont{H.}~\bibnamefont{Xiao}},
  \bibinfo{author}{\bibfnamefont{G.~F.} \bibnamefont{Chen}},
  \bibinfo{author}{\bibfnamefont{A.~J.} \bibnamefont{Taylor}},
  \bibnamefont{et~al.}, \bibinfo{journal}{Phys. Rev. B}
  \textbf{\bibinfo{volume}{93}}, \bibinfo{pages}{121110}
  (\bibinfo{year}{2016}).

\bibitem[{\citenamefont{Shao et~al.}(2018)\citenamefont{Shao, Sun, Wang, Xu,
  Sankar, Breindel, Cao, Fogler, Chou, Li et~al.}}]{Shao2018}
\bibinfo{author}{\bibfnamefont{Y.}~\bibnamefont{Shao}},
  \bibinfo{author}{\bibfnamefont{Z.}~\bibnamefont{Sun}},
  \bibinfo{author}{\bibfnamefont{Y.}~\bibnamefont{Wang}},
  \bibinfo{author}{\bibfnamefont{C.}~\bibnamefont{Xu}},
  \bibinfo{author}{\bibfnamefont{R.}~\bibnamefont{Sankar}},
  \bibinfo{author}{\bibfnamefont{A.~J.} \bibnamefont{Breindel}},
  \bibinfo{author}{\bibfnamefont{C.}~\bibnamefont{Cao}},
  \bibinfo{author}{\bibfnamefont{M.~M.} \bibnamefont{Fogler}},
  \bibinfo{author}{\bibfnamefont{F.}~\bibnamefont{Chou}},
  \bibinfo{author}{\bibfnamefont{Z.}~\bibnamefont{Li}}, \bibnamefont{et~al.}
  (\bibinfo{year}{2018}), \eprint{arXiv: 1806.01996}.

\bibitem[{\citenamefont{Ahn et~al.}(2017)\citenamefont{Ahn, Mele, and
  Min}}]{Ahn2017}
\bibinfo{author}{\bibfnamefont{S.}~\bibnamefont{Ahn}},
  \bibinfo{author}{\bibfnamefont{E.~J.} \bibnamefont{Mele}}, \bibnamefont{and}
  \bibinfo{author}{\bibfnamefont{H.}~\bibnamefont{Min}},
  \bibinfo{journal}{Phys. Rev. Lett.} \textbf{\bibinfo{volume}{119}},
  \bibinfo{pages}{147402} (\bibinfo{year}{2017}).

\bibitem[{\citenamefont{Tabert and Carbotte}(2016)}]{Tabert2016}
\bibinfo{author}{\bibfnamefont{C.~J.} \bibnamefont{Tabert}} \bibnamefont{and}
  \bibinfo{author}{\bibfnamefont{J.~P.} \bibnamefont{Carbotte}},
  \bibinfo{journal}{Phys. Rev. B} \textbf{\bibinfo{volume}{93}},
  \bibinfo{pages}{085442} (\bibinfo{year}{2016}).

\bibitem[{\citenamefont{Kresse and Hafner}(1993)}]{Kresse1993}
\bibinfo{author}{\bibfnamefont{G.}~\bibnamefont{Kresse}} \bibnamefont{and}
  \bibinfo{author}{\bibfnamefont{J.}~\bibnamefont{Hafner}},
  \bibinfo{journal}{Phys. Rev. B} \textbf{\bibinfo{volume}{47}}
  (\bibinfo{year}{1993}).

\bibitem[{\citenamefont{Kresse and Furthm{\"{u}}ller}(1996)}]{Kresse1996}
\bibinfo{author}{\bibfnamefont{G.}~\bibnamefont{Kresse}} \bibnamefont{and}
  \bibinfo{author}{\bibfnamefont{J.}~\bibnamefont{Furthm{\"{u}}ller}},
  \bibinfo{journal}{Computational Materials Science}
  \textbf{\bibinfo{volume}{6}}, \bibinfo{pages}{15} (\bibinfo{year}{1996}),
  \eprint{0927-0256(96)00008}.

\bibitem[{\citenamefont{Kresse and Furthm\"uller}(1996)}]{Kresse1996-2}
\bibinfo{author}{\bibfnamefont{G.}~\bibnamefont{Kresse}} \bibnamefont{and}
  \bibinfo{author}{\bibfnamefont{J.}~\bibnamefont{Furthm\"uller}},
  \bibinfo{journal}{Phys. Rev. B} \textbf{\bibinfo{volume}{54}},
  \bibinfo{pages}{11169} (\bibinfo{year}{1996}).

\bibitem[{\citenamefont{Perdew et~al.}(1996)\citenamefont{Perdew, Burke, and
  Ernzerhof}}]{John1996}
\bibinfo{author}{\bibfnamefont{J.~P.} \bibnamefont{Perdew}},
  \bibinfo{author}{\bibfnamefont{K.}~\bibnamefont{Burke}}, \bibnamefont{and}
  \bibinfo{author}{\bibfnamefont{M.}~\bibnamefont{Ernzerhof}},
  \bibinfo{journal}{Phys. Rev. Lett.} \textbf{\bibinfo{volume}{77}},
  \bibinfo{pages}{3865} (\bibinfo{year}{1996}).

\bibitem[{\citenamefont{He et~al.}(2014)\citenamefont{He, Xi, and Ku}}]{He2014}
\bibinfo{author}{\bibfnamefont{X.-G.} \bibnamefont{He}},
  \bibinfo{author}{\bibfnamefont{X.}~\bibnamefont{Xi}}, \bibnamefont{and}
  \bibinfo{author}{\bibfnamefont{W.}~\bibnamefont{Ku}} (\bibinfo{year}{2014}),
  \eprint{arXiv: 1410.2885}.

\bibitem[{\citenamefont{Preier}(1979)}]{Preier1979}
\bibinfo{author}{\bibfnamefont{H.}~\bibnamefont{Preier}},
  \bibinfo{journal}{Appl. Phys.} \textbf{\bibinfo{volume}{20}},
  \bibinfo{pages}{189} (\bibinfo{year}{1979}).

\bibitem[{\citenamefont{K\'ezsm\'arki et~al.}(2008)\citenamefont{K\'ezsm\'arki,
  Tomioka, Miyasaka, Demk\'o, Okimoto, and Tokura}}]{Tokura2008}
\bibinfo{author}{\bibfnamefont{I.}~\bibnamefont{K\'ezsm\'arki}},
  \bibinfo{author}{\bibfnamefont{Y.}~\bibnamefont{Tomioka}},
  \bibinfo{author}{\bibfnamefont{S.}~\bibnamefont{Miyasaka}},
  \bibinfo{author}{\bibfnamefont{L.}~\bibnamefont{Demk\'o}},
  \bibinfo{author}{\bibfnamefont{Y.}~\bibnamefont{Okimoto}}, \bibnamefont{and}
  \bibinfo{author}{\bibfnamefont{Y.}~\bibnamefont{Tokura}},
  \bibinfo{journal}{Phys. Rev. B} \textbf{\bibinfo{volume}{77}},
  \bibinfo{pages}{075117} (\bibinfo{year}{2008}).

\bibitem[{\citenamefont{Gu et~al.}(2014)\citenamefont{Gu, Wang, Zhou, Li, Cao,
  Zhang, Zhang, Gong, and Wang}}]{Gu2014}
\bibinfo{author}{\bibfnamefont{Y.}~\bibnamefont{Gu}},
  \bibinfo{author}{\bibfnamefont{K.}~\bibnamefont{Wang}},
  \bibinfo{author}{\bibfnamefont{H.}~\bibnamefont{Zhou}},
  \bibinfo{author}{\bibfnamefont{Y.}~\bibnamefont{Li}},
  \bibinfo{author}{\bibfnamefont{C.}~\bibnamefont{Cao}},
  \bibinfo{author}{\bibfnamefont{L.}~\bibnamefont{Zhang}},
  \bibinfo{author}{\bibfnamefont{Y.}~\bibnamefont{Zhang}},
  \bibinfo{author}{\bibfnamefont{Q.}~\bibnamefont{Gong}}, \bibnamefont{and}
  \bibinfo{author}{\bibfnamefont{S.}~\bibnamefont{Wang}},
  \bibinfo{journal}{Nanoscale Research Letters} \textbf{\bibinfo{volume}{9}},
  \bibinfo{pages}{24} (\bibinfo{year}{2014}), ISSN \bibinfo{issn}{1556-276X}.

\bibitem[{\citenamefont{Chen et~al.}(2017)\citenamefont{Chen, Chen, Zhong,
  Schneeloch, Zhang, Huang, Qu, Yu, Li, Gu et~al.}}]{Chen2017}
\bibinfo{author}{\bibfnamefont{Z.-G.} \bibnamefont{Chen}},
  \bibinfo{author}{\bibfnamefont{R.~Y.} \bibnamefont{Chen}},
  \bibinfo{author}{\bibfnamefont{R.~D.} \bibnamefont{Zhong}},
  \bibinfo{author}{\bibfnamefont{J.}~\bibnamefont{Schneeloch}},
  \bibinfo{author}{\bibfnamefont{C.}~\bibnamefont{Zhang}},
  \bibinfo{author}{\bibfnamefont{Y.}~\bibnamefont{Huang}},
  \bibinfo{author}{\bibfnamefont{F.}~\bibnamefont{Qu}},
  \bibinfo{author}{\bibfnamefont{R.}~\bibnamefont{Yu}},
  \bibinfo{author}{\bibfnamefont{Q.}~\bibnamefont{Li}},
  \bibinfo{author}{\bibfnamefont{G.~D.} \bibnamefont{Gu}},
  \bibnamefont{et~al.}, \bibinfo{journal}{Proceedings of the National Academy
  of Sciences} \textbf{\bibinfo{volume}{114}}, \bibinfo{pages}{816}
  (\bibinfo{year}{2017}), ISSN \bibinfo{issn}{0027-8424}.

\bibitem[{\citenamefont{Varshni}(1967)}]{Varshni1967}
\bibinfo{author}{\bibfnamefont{Y.}~\bibnamefont{Varshni}},
  \bibinfo{journal}{Physica} \textbf{\bibinfo{volume}{34}},
  \bibinfo{pages}{149} (\bibinfo{year}{1967}).

\bibitem[{\citenamefont{Wu et~al.}(2015)\citenamefont{Wu, Qin, Liang, Le, Fan,
  and Hu}}]{Wu2015}
\bibinfo{author}{\bibfnamefont{X.}~\bibnamefont{Wu}},
  \bibinfo{author}{\bibfnamefont{S.}~\bibnamefont{Qin}},
  \bibinfo{author}{\bibfnamefont{Y.}~\bibnamefont{Liang}},
  \bibinfo{author}{\bibfnamefont{C.}~\bibnamefont{Le}},
  \bibinfo{author}{\bibfnamefont{H.}~\bibnamefont{Fan}}, \bibnamefont{and}
  \bibinfo{author}{\bibfnamefont{J.}~\bibnamefont{Hu}}, \bibinfo{journal}{Phys.
  Rev. B} \textbf{\bibinfo{volume}{91}}, \bibinfo{pages}{081111}
  (\bibinfo{year}{2015}).

\end{thebibliography}

\end{document}